\documentclass[showpacs,pre]{revtex4}
\usepackage{graphicx}
\begin{document}
\title{Symmetry breaking of vortex patterns in a rotating harmonic potential}
\author{H.~Sakaguchi and H.~Takeshita}
\affiliation{Department of Applied Science for Electronics and Materials,
Interdisciplinary Graduate School of Engineering Sciences, Kyushu
University, Kasuga, Fukuoka 816-8580, Japan
}
\begin{abstract}
We numerically study the symmetry breaking instabilities of vortex patterns in a rotating harmonic potential using a type of Ginzburg-Landau equation. 
The configurations of vortex lattices change markedly by the symmetry-breaking instabilities, and then, some vortices move away from the confinement potential, which leads to the annihilation of vortices. 
The symmetry-breaking instabilities and the instabilities of vortex nucleation determine the parameter region of stable vortex patterns. We verify that the symmetry-breaking instabilities also occur in a type of complex Ginzburg-Landau equation.
\end{abstract}
\maketitle
\section{Introduction and model equation}
 Bose-Einstein condensates in trapped atomic gases at ultra-low temperatures have been intensively studied since their first experimental observation in 1995.~\cite{rf:1,rf:2,rf:3} In the Bose-Einstein condensates, quantized vortices and vortex lattices were found experimentally in 1999.~\cite{rf:4,rf:5} These vortex states have also been intensively studied theoretically.~\cite{rf:6,rf:7,rf:8}
 
The dynamics of the Bose-Einstein condensates in a rotating potential is described fairly well by the Gross-Pitaevskii (GP) equation: 
\begin{equation}
i\hbar\frac{\partial \phi}{\partial t}=\left [-\frac{\hbar^2}{2m}\nabla^2+g|\phi|^2+U({\bf r})-\Omega L_z\right ]\phi,
\end{equation}
where $m$ is the atomic mass, $U=1/kr^2$ is the harmonic potential for the trapping, $g=4\pi\hbar^2a/m$ is a coupling constant characterized by the s-wave scattering length $a$, $\Omega$ is the rotating frequency, and $L_z=-i\hbar(x\partial_y-y\partial_x)$. In this paper, we assume that $\hbar=1$ and $m=1$ by performing a scale transformation, and we consider only a two-dimensional system. 
The vortex states can be stabilized by the rotation term $-\Omega L_z\phi$. 
The solution with the lowest energy was approximately evaluated on the basis of the GP equation by Butts and Rokhsar.~\cite{rf:9} 
The linear perturbation $\delta \phi$ around a stationary solution $\phi$ obeys 
\begin{equation}
i\frac{\partial \delta \phi}{\partial t}=\left [-\frac{1}{2}\nabla^2+2g|\phi|^2\delta\phi+U({\bf r})-\Omega L_z\right ]\delta\phi+\phi^2\delta\phi^*.
\end{equation}
If $\delta\phi$ is expressed as $\delta\phi=(u_je^{-i\omega_jt}-v_j^*e^{i\omega_jt})e^{-i\mu t}$ and is substituted into eq.~(2), the Bogoliubov-de Gennes equation is obtained as
\begin{eqnarray}
\omega_ju_j&=&\left [-\frac{1}{2}\nabla^2+2g|\phi|^2\delta\phi+U({\bf r})-\Omega L_z\right ]u_j-\phi^2v_j,\nonumber\\
\omega_jv_j&=&-\left [-\frac{1}{2}\nabla^2+2g|\phi|^2\delta\phi+U({\bf r})-\Omega L_z\right ]v_j+\phi^{*2}u_j,
\end{eqnarray}
where $j$ denotes the mode number and $\omega_j$ is the eigenfrequency of the $j$th mode. The excitation of the surface mode is related to the nucleation of vortices.~\cite{rf:10} 
The excitation of the deformation mode of the vortex lattice is called the Tkachenko oscillation.~\cite{rf:11}

The energy and the total atomic number $M=\int|\phi|^2d{\bf r}$ are conserved in the time evolution of the GP equation. Therefore, the lowest energy state is not attained by the direct numerical simulation of the GP equation from general initial conditions. Tsubota et al. performed a numerical simulation of  the nucleation of vortices and the formation of a vortex lattice by introducing a damping term to the GP equation as~\cite{rf:12}
\begin{equation}
(i-\gamma)\frac{\partial \phi}{\partial t}=\left [-\frac{1}{2}\nabla^2+g|\phi|^2+U({\bf r})-\mu(t)-\Omega L_z\right ]\phi,
\end{equation}
where $\mu$ is the chemical potential and $\gamma$ is a damping parameter. 
The total atomic number $M=\int|\phi|^2d{\bf r}$ is not conserved in this system, but they continuously adjusted the chemical potential $\mu(t)$ so as to preserve the total number of condensates. 

In this paper, we study a type of Ginzburg-Landau equation to find stable vortex patterns. 
The model equation is written as
\begin{eqnarray}
\frac{\partial \phi}{\partial t}&=&\left [\frac{1}{2}\nabla^2-g|\phi|^2-U({\bf r})+\Omega L_z+\alpha(M_1(t)-M(t))\right ]\phi,\nonumber\\
\frac{dM_1}{dt}&=&\beta(M_0-M(t)),
\end{eqnarray}
where $M(t)=\int|\phi|^2d{\bf r}$, and $M_0$ is a parameter representing the target total atomic number, $M_1(t)$ is an additional variable, and $\alpha$ and $\beta$ are additional parameters.  The Ginzburg-Landau equation is obtained from eq.~(4) by retaining only the damping term in the left-hand side of eq.~(4) and setting the damping parameter to $\gamma=1$.
The second equation for $M_1(t)$ acts as negative feedback to keep the total atomic number $M(t)$ close to $M_0$. 
This model equation is an efficient model for obtaining stationary states with lower energy for a certain fixed value $M_0$ of the atomic number, because the Ginzburg-Landau equation is a variational system. 
The Ginzburg-Landau-type equation eq.~(5) can be directly obtained from the GP equation eq.~(1) by a change of variable $t\rightarrow it$, in which $t$ is interpreted as the imaginary time. However, in this paper, we consider eq.~(5) to be  a limit of eq.~(4) for large $\gamma$ with the timescale changed from $\gamma^{-1} t$ to $t$.  In this interpretation, $t$ is the actual time. An intermediate model  between eq.~(1) and eq.~(5), which has a similar form to eq.~(4), will be studied in \S 4. The stationary solutions to eq.~(5) are  the stationary solutions to the GP equation satisfying $M(t)=M_0$. These solutions are also stationary solutions to eq.~(4) if the chemical potential $\mu$ is suitably adjusted. 
The chemical potential $\mu$ is expressed as $\mu=\alpha(M_1(t)-M_0)$ for the stationary solutions to eq.~(5).  We have used this type of Ginzburg-Landau equation in a previous paper to find soliton solutions to the GP equation for attractive interaction.~\cite{rf:13} In numerical simulations in this paper, the parameter values of $M_0=210$, $\alpha=3$, and $\beta=0.5$ are used and a harmonic potential $U({\bf r})=1/2kr^2$ with $k=0.25$ is assumed.  We performed numerical simulations by the split-step Fourier method with $256\times 256$ modes. 

The linear stability of the stationary solutions $\phi$ for eq.~(5) can be studied using the linear equation for the perturbation $\delta \phi$: 
\begin{equation}
\frac{\partial \delta \phi}{\partial t}=\left [\frac{1}{2}\nabla^2-2g|\phi|^2\delta\phi-U({\bf r})+\mu+\Omega L_z\right ]\delta\phi-\phi^2\delta\phi^*,
\end{equation}
if $\alpha(M_1-M)=\mu=$ const.  Various types of linear instability can occur in the Ginzburg-Landau-type equation when the eigenvalue becomes positive. The positive eigenvalue corresponds to the situation where the eigenfrequency $\omega_j$ is negative in eq.~(3). When a surface mode exhibits an instability, new vortices are nucleated. When an instability occurs in a deformation mode of a vortex lattice, the configuration of the vortex lattice changes markedly. This is the 
main subject of this paper. 
\section{Symmetry Breaking in Point-Vortex Systems}
Campbell and Ziff studied various vortex patterns in a rotating superfluid confined in a cylinder of radius $R$ using the energy of point vortices and their mirror images.~\cite{rf:14} They showed that many vortex patterns with various numbers of vortices are stable at the same frequency $\Omega$, and that multiple configurations are possible even for the same vortex number when it is larger than 9. That is, there are many vortex patterns, that are local minima of the energy.
We explain symmetry-breaking instabilities in the point vortex systems first, which will help with the understanding of the results of direct numerical simulations using eq.~(5) described in \S 3.
\begin{figure}[tbp]
\includegraphics[width=13cm]{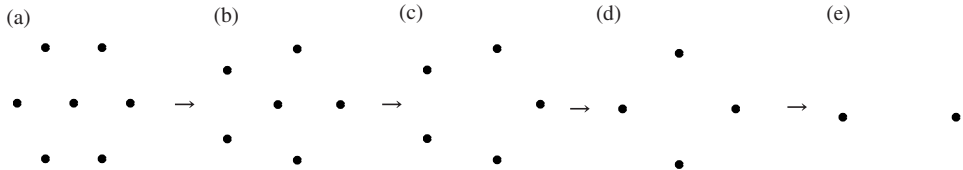}
\caption{Successive vortex patterns including (a) seven vortices, (b) six vortices, (c) five vortices, (d) four vortices, and (e) two vortices.}
\label{fig1}
\end{figure}
The energy $f$ of a system of $N$ point vortices in a rotating vessel with radius $R$ and frequency $\Omega$ is expressed as,
\begin{equation}
f=-\sum_{j>i=1}^N\ln r_{i,j}+(1/2)\sum_{i,j=1}^N\ln r_{i,j}^{\prime}+(1/2)\Omega\sum_{i=1}^N r_i^2,
\end{equation}
where $r_i=\sqrt{x_i^2+y_i^2}$ is the distance between the center and the $i$th vortex located at $(x_i,y_i)$, $ r_{i,j}=\sqrt{(x_i-x_j)^2+(y_i-y_j)^2}$ is the distance between the $i$th and $j$th vortices, $r_{i,j}^{\prime}=\sqrt{(x_i-R^2x_j/r_j^2)^2+(y_i-R^2y_j/r_j^2)^2}$ is the distance between the $i$th vortex and the mirror image of the $j$th vortex located at $(R^2x_j/r_j^2,R^2y_j/r_j^2)$, and $\Omega$ is the external frequency.  The equation of motion for the $i$th vortex is determined as 
\begin{equation}
\frac{dx_i}{dt}=\frac{\partial f}{\partial y_i}, \frac{dy_i}{dt}=-\frac{\partial f}{\partial x_i}.
\end{equation}

To efficiently find stable stationary solutions satisfying $dx_i/dt=dy_i/dt=0$, we study a variational system:
\begin{equation}
\frac{dx_i}{dt}=-\frac{\partial f}{\partial x_i}, \frac{dy_i}{dt}=-\frac{\partial f}{\partial y_i}.
\end{equation}
This model equation corresponds to the Ginzburg-Landau-type equation eq.~(5) for the GP equation.  The direction of the velocity of a point vortex in the variational model is rotated by $-\pi/2$ from that of the velocity given by eq.~(8). We can study stable configurations of point vortices and their instabilities by numerical simulations using eq.~(9). 

Firstly, we consider a single vortex located at $(r,0)$. The position of the mirror image of the single vortex is  $(R^2/r,0)$. The energy $f$ of the point vortex and the velocity $v_{r}$ are calculated as 
\begin{equation}
f=(1/2)\ln(R^2-r^2)+(1/2)\Omega r^2,\;\; v_{r}=1/(R^2/r-r)-\Omega r.   
\end{equation} 
(The angular velocity $v_{\theta}$ obtained from eq.~(8) is expressed as $v_{\theta}=1/(R^2/r^2-r)-\Omega r$.)
The energy of the single vortex is $f=\ln R$ at $r=0$ and $f=-\infty$ at $r=R$. The origin $r=0$ becomes a local minimum for $\Omega>1/R^2$.
That is, a single vortex becomes a stable state for $\Omega>1/R^2$. 
This instability for $\Omega<1/R^2$ is a symmetry-breaking instability for a single vortex, because the most symmetrical position $r=0$ becomes an unstable position. 
We denote the critical values of $\Omega$ for the symmetry-breaking instability 
 obtained from eq.~(9) including $N$ point vortices as $\Omega_{cN,0}$. In our system confined in a harmonic potential $U=(1/2)kr^2$, the effective radius $R$ can be estimated as $R=(4M_0/\pi k)^{1/4}$  by the Thomas-Fermi approximation, where $\phi$ becomes zero. 
In our model system with $M_0=210$ and $k=0.25$, the effective radius $R$ and the critical value are evaluated as $R=5.71$ and $\Omega_{c1,0}\sim 0.0306$. In the time evolution of eq.~(9), a single vortex stays at the center for $\Omega\ge\Omega_{c1,0}$, but it becomes unstable and the vortex moves away from the center for $\Omega<\Omega_{c1,0}$. This instability is due to the force from the mirror image. 

Next we consider two vortices located at $(r_1,0)$ and $(-r_2,0)$. The vortex position in the case of a symmetrical solution satisfying $r_1=r_2=d/2$ is obtained as a solution of $\Omega=8d^2/(16R^4-d^4)+2/d^2$, which is approximated to $d=\sqrt{2/\Omega}$, if the mirror-image force is neglected. The stationary solution $r_1=r_2=d/2$ disappears for $\Omega<0.065$ when $R=5.71$. However, a symmetry-breaking instability occurs at a larger critical value of $\Omega_{c2,0}=0.085$ in the time evolution of eq.~(9). For $\Omega<0.085$, one vortex moves to the center and the other one moves towards the wall at $r=R$ owing to the symmetry-breaking instability. The symmetry breaking implies that the energy $f$ is not a local minimum at $r_1=r_2=d/2$ for $\Omega<\Omega_{c2,0}$.   The symmetry-breaking instability does not occur if the image force is neglected. In the time evolution of eq.~(9), the creation and annihilation of point vortices do not occur, and the distance $r$ from the center of one vortex can become larger than $R$; in this case, eq.~(9) has no physical meaning.

Similar symmetry-breaking instabilities also take place for systems including a large number of vortices. Figure 1 displays vortex patterns that are expected to appear after successive symmetry-breaking instabilities from the first pattern, which has 7 vortices. This sequence of vortex patterns was obtained by a numerical simulation using eq.~(9) by gradually decreasing the parameter $\Omega$. In  the actual numerical simulation, the variables $x_j$ and $y_j$ for the $j$th vortex were removed when $r_j$ reached $R$. This is because the vortex is expected to disappear at the wall, where $r=R$. We have numerically confirmed  that these successive symmetry-breaking instabilities do not occur if the forces from the mirror images are neglected.  
 The critical values of the successive symmetry-breaking instabilities were numerically found to be $\Omega_{c7,0}=0.231, \Omega_{c6,0}=0.209,\Omega_{c5,0}=0.181$, and $\Omega_{c4,0}=0.143$ for $R=5.71$.  
We can evaluate another value of the effective radius, $R=7.5$, which was found by a direct numerical simulation of eq.~(5) as a radius satisfying $|\phi(r)|\sim 0$ for $r>R$. If the cylinder radius is assumed to be $R=7.5$, the critical values of the successive symmetry-breaking instabilities are $\Omega_{c7,0}=0.133, \Omega_{c6,0}=0.116,\Omega_{c5,0}=0.1$, $\Omega_{c4,0}=0.083$, $\Omega_{c2,0}=0.049$, and $\Omega_{c1,0}=0.017$. The critical values  depend strongly on the radius $R$, because the mirror force is essential for the symmetry-breaking instabilities.  The mirror symmetry is broken at the instability of vortex patterns (a), (b), (c), and (e). On the other hand, for the vortex pattern (d) with $N=4$, the square configuration becomes a rhombic pattern at the symmetry-breaking instability. After the instability, two vortices move outwards and disappear at $r=R$  and two vortices survive, as shown in Fig.~1(e). 

\begin{figure}[tbp]
\includegraphics[height=4.5cm]{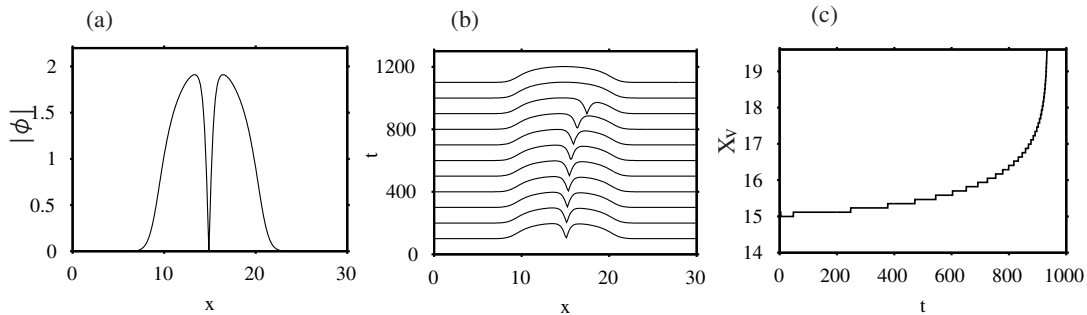}
\caption{(a) Profile $|\phi|$ at the section $y=L/2$ for $\Omega=0.14$. (b) Time evolution of $|\phi|$ at the section $y=L/2$ for $\Omega=0.12$. (c) Time evolution of the vortex position $X_v$ for $\Omega=0.12$.}
\label{fig2}
\end{figure}
\section{Symmetry Breaking and Vortex Annihilation in the Ginzburg-Landau-Type 
Equation}
In this section, we study vortex patterns and their symmetry-breaking instabilities by numerical simulation using eq.~(5) with $g=1$. The system size is $L\times L=30\times 30$. The center of the harmonic potential is assumed to be $(L/2,L/2)$. In contrast to the point-vortex system, our vortices have a finite core size. Our vortices are not confined by a cylindrical wall but by a harmonic potential. Also, the nucleation and annihilation of vortices occur naturally in this system in contrast to the point-vortex system in \S 2. 

We first performed the numerical simulation of a single vortex, which was initially set near the center.  Figure 2(a) displays $|\phi|$ for $\Omega=0.14$ in the cross section at $y=L/2$. A vortex is stably located at the center for the frequency $\Omega=0.14$. The modulus $|\phi|$ is 0 at the position of the vortex, and $\phi$ is also almost 0 for $r>R\sim 7.5 $ owing to the confinement by the harmonic potential. Figure 2(b) displays the time evolution of $|\phi(x,y)|$ at the section $y=L/2$ by eq.~(5), and Fig.~2(c) displays the time evolution of the vortex position when $\Omega=0.12$. The single-vortex state is unstable at this frequency, and the vortex moves away from the center and finally disappears. We denote the critical values of $\Omega$ for the symmetry-breaking instability in eq.~(5) including $N$ vortices as $\Omega_{cN}$. 
 The critical value is $\Omega_{c1}\sim 0.13$, which is interpreted to correspond to $\Omega_{c1,0}=0.0306$ for $R=5.71$ or $\Omega_{c1,0}=0.017$ for $R=7.5$ in the point-vortex theory. 
On the other hand,  new vortices are nucleated from the instability of the surface mode at $\Omega=0.31$. We denote the critical values of $\Omega$ for the vortex nucleation in eq.~(5) including $N$ vortices as $\Omega_{nN}$. That is, $\Omega_{n1}=0.31$.
\begin{figure}[tbp]
\includegraphics[height=4.5cm]{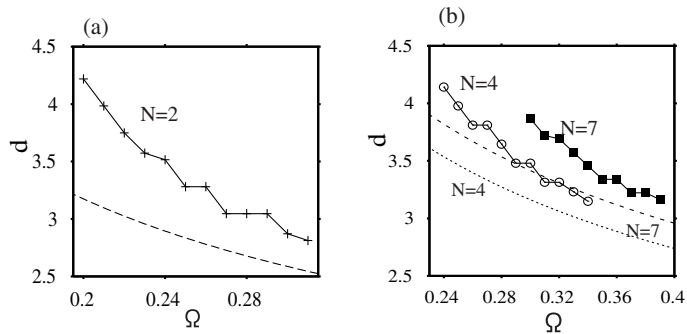}
\caption{(a) Distance $d$ between two vortices for $N=2$. (b) Distances $d$ between two vortices for $N=4$ and $N=7$.
}
\label{fig3}
\end{figure}
\begin{figure}[tbp]
\includegraphics[height=5.2cm]{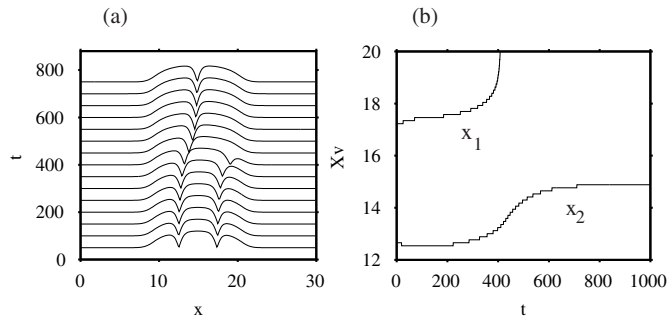}
\caption{(a) Time evolution of $|\phi|$ at the section $y=L/2$ for $\Omega=0.18$. (b) Time evolution of the two vortex centers $x_1$ and $x_2$ for $\Omega=0.18$. 
}
\label{fig4}
\end{figure}

Next, we consider a system including two vortices, which are initially located at $(x_1,L/2)$ ($x_1>L/2$) and $(x_2,L/2)$ ($x_2<L/2$).  
Figure 3(a) displays the distance $d=x_1-x_2$ between the two vortices in the stationary solution as a function of $\Omega$ obtained by a long time numerical simulation using eq.~(5). The dashed curve is the theoretical curve estimated from  $\Omega=8d^2/(16R^4-d^4)+2/d^2$ by the point-vortex theory. Here, $R=7.5$ is used; however, the effect of the mirror image is very small and the curve is well approximated at $d=\sqrt{2/\Omega}$ for the parameter range $0.2<\Omega<0.31$. 
 In the numerical simulation of eq.~(5), the symmetry-breaking instability occurs at $\Omega_{c2}=0.19$, which corresponds to $\Omega_{c2,0}=0.085$ for $R=5.71$ or $\Omega_{c2,0}=0.049$ for $R=7.5$ in the point-vortex theory. Although similar types of symmetry breaking occurs according to eq.~(5) and eq.~(9), both the distance between the two vortices and the critical value of the symmetry breaking are  quantitatively different from those in the point-vortex theory.
Figure 4(a) displays the time evolution of $|\phi|$ at $y=L/2$ for  $\Omega=0.18$. Figure 4(b) displays the time evolutions of the $x$-coordinates of the left and right vortices.
It is clearly seen that one of the vortices moves to the center and the other one moves outwards and disappears. A single-vortex state appears after a long time, which is located at the center.  On the other hand, the vortex nucleation occurs at $\Omega_{n2}=0.32$ for this system with $N=2$. 

We have studied a system including four vortices forming a square and a system including seven vortices forming a triangular lattice, whose configurations are respectively similar to Figs.~1(d) and 1(a). Figure 3(b) displays the distances between the nearest-neighbor vortices for the systems with $N=4$ and $N=7$. 
The dashed curves are theoretically estimated  by the point-vortex theory, where $R=7.5$ is used. 
The theoretically estimated values are smaller than the numerical results.
The symmetry-breaking instability occurs at $\Omega_{c4}=0.23$ for the system with $N=4$ and at $\Omega_{c7}=0.29$ for the system with $N=7$. 
Vortex nucleation occurs at $\Omega_{n4}=0.35$ for the system with $N=4$ and at $\Omega_{n7}=0.40$ for the system with $N=7$.
The critical value $\Omega_{nN}$ of the vortex nucleation increases with the vortex number $N$. The vortex pattern with vortex number $N$ is stable in the parameter region between $\Omega_{cN}$ and $\Omega_{nN}$. The parameter regions mutually overlap for different values of $N$.  We can therefore conclude that multiple vortex patterns are also stable in our system for the same frequency $\Omega$ from the relation bwteen the critical values $\Omega_{cN}$ and $\Omega_{nN}$ for various $N$.

\begin{figure}[tbp]
\includegraphics[width=12cm]{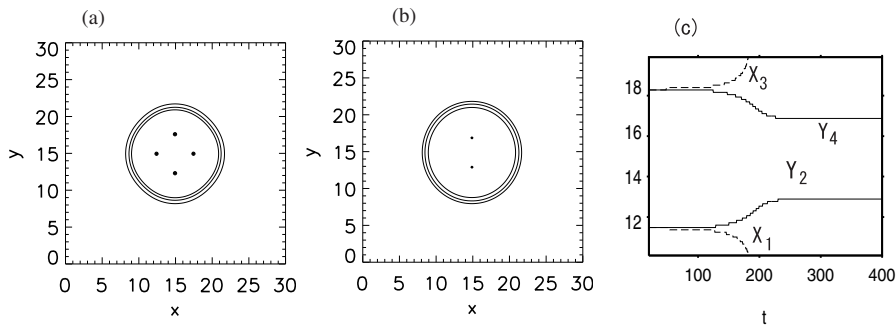}
\caption{(a) Vortex pattern at $t=100$ including four vortices when $\Omega=0.21$. (b) Vortex pattern at $t=400$ when $\Omega=0.21$. (c) Time evolutions of the $x$- and $y$-coordinates of the vortex positions.}
\label{fig5}
\end{figure}
\begin{figure}[tbp]
\includegraphics[height=5.5cm]{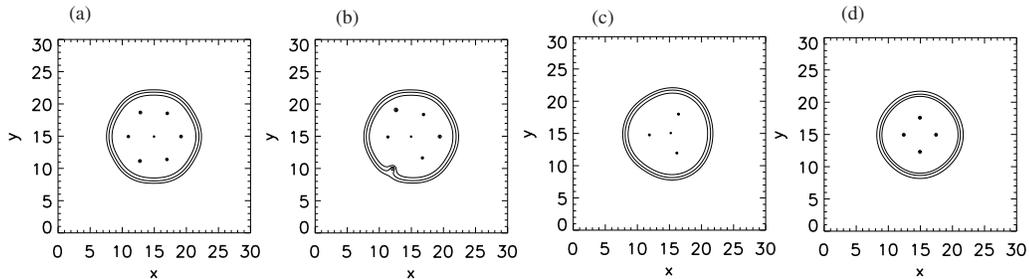}
\caption{Vortex patterns at (a) $t=100$, (b) $t=150$, (c) $t=1000$, and (d) $t=1750$ for a system with $N=7$ when $\Omega=0.28$.}
\label{fig6}
\end{figure}
Figures 5(a) and 5(b) display snapshot vortex patterns at $t=100$ and $t=400$ as a result of the symmetry-breaking instability when $\Omega=0.21$ for the system including four vortices initially located on the lines $x=L/2$ and $y=L/2$. The two vortices on the line $y=L/2$ move outwards, and the two vortices on the line $x=L/2$ survive, which is qualitatively consistent with the point-vortex theory. Figure 5(c) displays the time evolutions of the $x$-coordinates $X_1$ and $X_3$ of the vortex positions on the line $y=L/2$ and the $y$-coordinates $Y_2$ and $Y_4$ on the line $x=L/2$. The two vortices on the line $x=L/2$ approach each other and the other two vortices on the lines $y=L/2$ move away from each other.

Figures 6(a)-6(d) display four snapshots of vortex patterns at (a) $t=100$, (b) $t=150$, (c) $t=1000$, and (d) $t=1750$ for a system with $N=7$ when $\Omega=0.28$. As a result of the symmetry breaking, one vortex moves away at $t=150$, then the other two vortices also move away. Four vortices survive at $t=1000$; the configuration of the four vortices changes markedly and  they finally form a square pattern at $t=1750$.  
\begin{figure}[tbp]
\includegraphics[height=4.5cm]{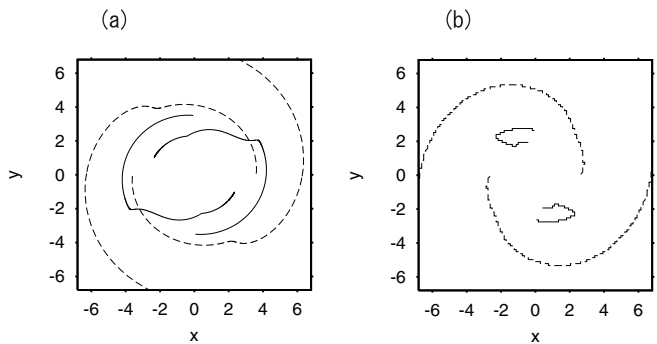}
\caption{(a) Trajectories of four vortices obtained from eq.~(11) when $\Omega=0.081, R=7.5$ and $\gamma=0.1$. 
(b) Trajectories of four vortices by the complex Ginzburg-Landau-type model eq.~(12) when $\Omega=0.21$ and $\gamma=0.1$. Solid curves correspond to the vortices taht survive and dashed curves correspond to the vortices that move outwards and disappear.}
\label{fig7}
\end{figure}

These numerical results show that symmetry-breaking instabilities also occur according to eq.~(5), which are qualitatively the same as those found in \S 2 for the point vortices. However, the critical values $\Omega_{cN,0}$ in the point-vortex system given by eq.~(9) and $\Omega_{cN}$ given by eq.~(5) are rather different. This is partly because the core size of the vortex is finite in the system given by eq.~(5), in contrast to the point vortices with infinitesimal core size. 
As the nonlinear parameter $g$ is increased, the core size decreases as $1/\sqrt{g}$ for large $g$ from an argument based on scaling. That is, if the core size is denoted as $\xi$, $(1/2)\nabla^2\phi\sim 1/(2\xi^2)\phi\sim g|\phi|^2\phi$ for large $g$ in eq.~(1) with $\hbar=m=1$. The point-vortex theory is therefore expected to give a better approximation in the case of large $g$. 
Furthermore, the harmonic potential is used in the system based on eq.~(5) to confine vortex patterns in contrast to the circular confinement by a cylindrical wall with radius $R$ for the system given by eq.~(9).  

\section{Symmetry Breaking in a Complex Ginzburg-Landau-Type Equation}
The symmetry-breaking instability does not occur in the GP equation eq.~(1) or in eq.~(8) for the point vortices, because there is no energy dissipation. If some dissipation terms are included, the symmetry-breaking instability is expected to occur. We have verified this using intermediate models between eq.~(8) and eq.~(9), and between eq.~(1) and eq.~(5). A model equation  for the point vortices with a small amount of dissipation is expressed as  
\begin{equation}
\frac{dx_i}{dt}=\gamma_0\frac{\partial f}{\partial y_i}-\gamma_1\frac{\partial f}{\partial x_i}, \frac{dy_i}{dt}=-\gamma_0\frac{\partial f}{\partial x_i}-\gamma_1\frac{\partial f}{\partial y_i},
\end{equation}
where $\gamma_0=1/\sqrt{1+\gamma^2}$, and $\gamma_1=\gamma/\sqrt{1+\gamma^2}$ using a parameter $\gamma$ representing the degree of dissipation. 
This model equation has the form of a linear combination of eqs.~(8) and (9). 
Similarly, we can construct a model equation from eqs.~(1) and (5) as 
\begin{eqnarray}
\frac{\partial \phi}{\partial t}&=&(-i\gamma_0+\gamma_1)\left [\frac{1}{2}\nabla^2-g|\phi|^2-U({\bf r})+\Omega L_z\right ]\phi+\gamma_1\alpha (M_1(t)-M(t))\phi,\nonumber\\
\frac{dM_1}{dt}&=&\beta(M_0-M(t)),
\end{eqnarray}
where $\gamma_0=1/\sqrt{1+\gamma^2}$, and $\gamma_1=\gamma/\sqrt{1+\gamma^2}$.
This is a model equation with a similar form to eq.~(4) and is interpreted as a type of complex Ginzburg-Landau equation. 
When $\gamma_0=1$ and $\gamma_1=0$, eq.~(11) is reduced to eq.~(8), and eq.~(12) reduced to eq.~(1) when $\hbar=m=1$. When $\gamma_0=0$ and $\gamma_1=1$, eq.~(11) is reduced to eq.~(9) and eq.~(12)  reduced to eq.~(5). 
We have performed a numerical simulation of vortex patterns with four vortices initially located on the lines of $x=L/2$ or $y=L/2$. 
Figure 7(a) displays the trajectories of the four vortices using the point-vortex model eq.~(11) when $\Omega=0.081$ obtained from $R=7.5$, and $\gamma=0.1$. Figure 7(b) displays similar trajectories of the four vortices in the complex Ginzburg-Landau-type equation eq.~(12) when $\Omega=0.21, \gamma=0.1$, and $g=1$. (For comparison with Fig.~7(a), the origin of the coordinates is shifted to $(0,0)$ in Fig.~7(b).)
The symmetry-breaking instability occurs for the vortex patterns and two vortices move outward and disappear. The other two vortices move inwards, and find new stationary stable positions. These symmetry-breaking instabilities are essentially the same as those obtained from eqs.~(5) and (9). However, the rotational motion (spiral-like motion) of vortices is observed in these models, although the motion is rather complicated. The rotational trajectories are characteristic of the motion of vortices given by the GP equation eq.~(1) and eq.~(8) for the point vortices.  
These numerical simulations suggest that the symmetry-breaking instabilities are not only found  in the imaginary models eqs.~(5) and (9), and the instabilities are realistic if some energy dissipation is involved in the system. Rotational or spiral-like trajectories seem to be more realistic than the straight trajectories shown in Figs.~4 and 5, and such trajectories might be found in experiments.  The straight trajectories numerically obtained by eq.~(5) and shown in Figs.~4 and 5 are expected to appear in the limit of strong dissipation.

\begin{figure}[tbp]
\includegraphics[height=6.cm]{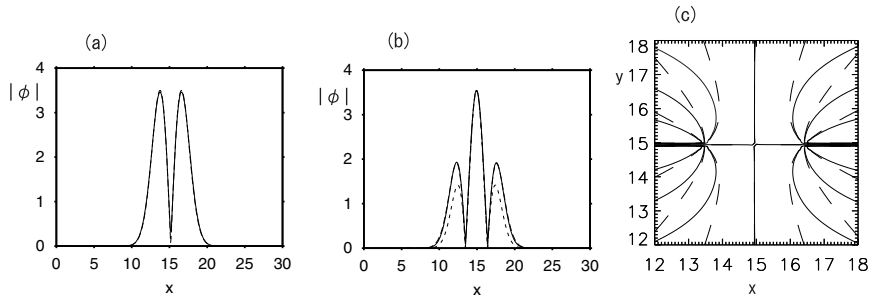}
\caption{(a) Profile $|\phi(x,y)|$ of a single-vortex pattern at $y=L/2$ for  $g=0.005,\Omega=0.49$, and $M_0=210$ using eq.~(5). The dashed curve denotes  $4.09|x-L/2|\exp\{-0.25(x-L/2)^2\}$, although it almost overlaps with the solid curve of $|\phi|$.
(b) Profile of $|\phi(x,L/2)|$ of a two-vortex pattern for $g=0.005,\Omega=0.49$, and $M_0=210$. The dashed curve denotes $1.63|(x-L/2)^2-1.47^2|\exp\{-0.205(x-L/2)^2\}$, and the dotted curve denotes $1.63|(x-L/2)^2-1.47^2|\exp\{-0.25(x-L/2)^2\}$. (c) Comparison of contour plots of the phase of $\phi$ obtained from eq.~(5) and the ansatz eq.~(16). The contour lines of $\phi=n\pi/4$ with $n=0,1,2,\cdots,8$ are drawn.}
\label{fig8}
\end{figure}
\section{Vortex Patterns  near the Lowest-Landau-Level Regime}
The GP equation eq.~(1) with $m=\hbar=1$  is expressed as 
\begin{equation}
i\frac{\partial \phi}{\partial t}=\frac{1}{2}\left [\left (i\frac{\partial}{\partial y}+\Omega x\right )^2+\left (i\frac{\partial}{\partial x}-\Omega y\right )^2 \right]\phi,
\end{equation}
in the special case when $g=0$ and $\Omega=\sqrt{k}$. There is a family of vortex solutions to eq.~(13):
\begin{equation}
\phi=\{x-x_0+i(y-y_0)\}\exp[-\Omega/2\{(x-x_0)^2+(y-y_0)^2\}]\exp[i\Omega (x_0y -y_0x)],
\end{equation}
where $(x_0,y_0)$ is the core position of a single vortex.  
This is equivalent to the wave function for an electron in a magnetic field at the lowest Landau level. 
We therefore call the parameter region of $g\sim 0$ and $\Omega\sim \sqrt{k}$ the lowest-Landau-Level regime (L-L-L regime). 
A large number of vortices are nucleated near $\Omega=\sqrt{k}$ if $g$ is not small. In the parameter region, Ho proposed a solution for a vortex pattern including $N$ vortices:
\begin{equation}
\phi\sim \Pi_{j=1}^N[(x-x_j)+i(y-y_j)]\exp\{-\Omega/2(x^2+y^2)\},
\end{equation}
where $(x_j,y_j)$ is the core position of the $j$th vortex.~\cite{rf:15} However, such a Gaussian behavior  was not observed for the envelope of $\phi$ with a large number of vortices. It is considered that the Thomas-Fermi form is a better approximation.~\cite{rf:16} The Thomas-Fermi approximation is suitable when $g$ is not small. We have studied vortex patterns with a few vortices in the L-L-L regime, because the wave function of a single vortex is explicitly given as eq.~(14), and a theoretical approach different from the point-vortex theory might be possible. 

We have performed numerical simulations using eq.~(5) by setting a few vortices as an initial condition for sufficiently small values of $g$ at $\Omega=0.49$, $k=0.25$ $\alpha=3$, $\beta=5$, and $M_0=210$. 
Figure 8(a) displays the profile of $|\phi(x,y)|$ at the section $y=L/2$ of a single vortex when $g=0.005$ and $\Omega=0.49$. The dashed curve denotes  $A|x-L/2|\exp\{-0.25(x-L/2)^2\}$,
where $A$ is calculated as $\sqrt{M/(4\pi)}\sim 4.09$ from the normalization condition. The vortex solution given by eq.~(14) is a good approximation. 

Figure 8(b) displays the profile of $|\phi|$ at the section of $y=L/2$ for a system including two vortices at $g=0.005$. The distance between the two vortices is 2.93 at $g=0.005$.
The dotted curve is the approximation $|\phi|=A|(x-L/2)^2-x_0^2|\exp\{-0.25(x-L/2)^2\}$, which is obtained from  the ansatz by Ho:
\begin{equation}
\phi=A\{x-L/2-x_0+i(y-L/2)\}\{x-L/2+x_0+i(y-L/2)\}\exp[-0.25\{(x-L/2)^2+(y-L/2)^2\}],
\end{equation}
and the vortex position $x_0=1.47$ obtained from the numerical simulation is used. The ansatz given by eq.~(16) is not satisfactory even for the small value of $g$. The dashed curve, which almost overlaps with the numerical result, is   $|\phi|=A|(x-L/2)^2-x_0^2|\exp\{-0.205(x-L/2)^2\}$. This Gaussian approximation is rather good.  Figure 8(c) displays a contour plot of the phase of $\phi$ given by tan$^{-1}({\rm Re}\phi/{\rm Im} \phi)$. The solid curve denotes the numerical results and the dashed curve is obtained from the ansatz eq.~(16). These figures imply that the ansatz is unsatisfactory, although a Gaussian approximation might be suitable. 

\begin{figure}[tbp]
\includegraphics[height=5.cm]{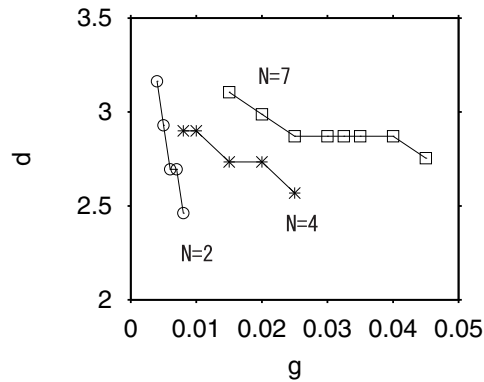}
\caption{Distances between the nearest-neighboring vortices in systems with $N=2,4$, and 7 as a function of $g$ for $\Omega=0.49$.}
\label{fig9}
\end{figure}
 We have studied the distance $d$ as a function of $g$ when $\Omega=0.49$ in the two-vortex, four-vortex, and seven-vortex systems studied in \S 3.  Figure 9 displays the distances between the nearest-neighboring vortices in systems with $N=2,4$, and 7.
The vortex patterns are stable only in small parameter regions with small values of $g$. The stable parameter region for $N=2$ is particularly small.  
Figure 9 suggests that the vortex number $N$ increases with $g$. 
When $g$ is increased beyond the critical value on the right, the vortex nucleation occurs from the surface-mode instability and the vortex number increases. When $g$ is decreased beyond the critical value on the left, vortices disappear as a result of the symmetry-breaking instabilities. The vortex patterns with different $N$ are multi stable in overlapping parameter regions. These behaviors are similar to the case of $g=1$ studied  in \S 3.  
However, we have not succeeded in evaluating the distance between the nearest-neighboring vortices theoretically, because a suitable ansatz for the wave function other than eqs.~(15) or (16) has not been found yet. Note that the $g$ dependence of distance $d$ does not appear in the point-vortex theory.

\section{Summary and Discussion}
We have proposed a type of Ginzburg-Landau equation. Using the model equation, we can efficiently find stable vortex patterns in rotating Bose-Einstein condensates. We have found that multiple-vortex patterns are stable at the same frequency $\Omega$ for $g=1$. We have also found that symmetry-breaking instabilities occur for vortex patterns, which lead the annihilation of vortices. 

The symmetry-breaking instabilities are qualitatively consistent with those obtained by the point-vortex theory, but the critical values are considerably different. 
It is left to a future study to estimate the critical values by taking the effects of the harmonic potential and the finite core size into consideration. 
The forces from the mirror images are essential for the symmetry-breaking instabilities in the point-vortex theory. 
From the analogy with the point-vortex system, we think that the confinement effect due to the harmonic potential is very important for causing the symmetry-breaking instabilities in the Ginzburg-Landau type equation. 

We have verified that symmetry-breaking instabilities also occur in a type of complex Ginzburg-Landau equation when the energy dissipation is small. Spiral-like trajectories have been numerically obtained, although the trajectories are rather complicated. We expect that such spiral-like trajectories of vortices might be observed in real experiments. 

The quantitative agreement between the point-vortex theory and the numerical simulation at $g=1$ was not satisfactory. 
In the limit of large $g$, the core size of a vortex becomes infinitesimal, because the core size is scaled as $1/\sqrt{g}$ for large $g$ from an argument based on the scale transformation. This is intuitively because the stronger repulsive interaction for larger $g$ makes the density $|\phi|^2$ more uniform and therefore the core size smaller. The point-vortex approximation is therefore expected to improve in the case of large $g$. We have also performed a numerical simulation near the L-L-L regime, where $\Omega\sim \sqrt{k}$ and  $g$ is sufficiently small in contrast to the limit of the point-vortex approximation when $g\rightarrow +\infty$.   We have found stable vortex patterns with $N=2,4$, and 7, when $g$ is sufficiently small. The single vortex is well approximated by eq.~(14); however, the ansatz eq.~(16) is not satisfactory for a two-vortex system. It is left to future studies to find a better approximation of $\phi$ in this parameter range and obtain a theoretical value for the distances among vortices.

\end{document}